\documentclass{article}
\usepackage[totalwidth=13.0cm,totalheight=20.0cm]{geometry}

\usepackage{latexsym,amsthm,amsmath,amssymb,url}

\newtheorem{lemma}{Lemma}
\newtheorem{corollary}{Corollary}

\newtheorem{theorem}{Theorem}

\newcommand{\NP}{\text{\normalfont NP}}
\renewcommand{\P}{\text{\normalfont P}}

\begin{document}

\title{Note on Maximal Bisection above Tight Lower Bound}

\author{
Gregory Gutin and Anders Yeo\\
\small Department of Computer Science\\[-3pt]
\small  Royal Holloway, University of London\\[-3pt]
\small Egham, Surrey TW20 0EX, UK\\[-3pt]
\small \texttt{gutin|anders@cs.rhul.ac.uk}
}
\date{}
\maketitle

\newenvironment{compress}{\baselineskip=10pt}{\par}
\begin{abstract}
\noindent
In a graph $G=(V,E)$, a bisection $(X,Y)$ is a partition of $V$ into sets $X$ and $Y$ such that $|X|\le |Y|\le |X|+1$. The size of $(X,Y)$ is the number of edges
between $X$ and $Y$. In the Max Bisection problem we are given a graph $G=(V,E)$ and are required to find a bisection of maximum size.
It is not hard to see that $\lceil |E|/2 \rceil$ is a tight lower bound on the maximum size of a bisection of $G$.

We study parameterized complexity of the following parameterized problem called Max Bisection above Tight Lower Bound (Max-Bisec-ATLB): decide whether a graph $G=(V,E)$ has a bisection of size at least $\lceil |E|/2 \rceil+k,$ where $k$ is the parameter. We show that this parameterized problem has a kernel with $O(k^2)$ vertices and $O(k^3)$ edges, i.e., every instance of Max-Bisec-ATLB is equivalent to an instance of Max-Bisec-ATLB on a graph with at most $O(k^2)$ vertices and $O(k^3)$ edges.
\end{abstract}

\section{Introduction}\label{section:intro}
In a graph $G=(V,E)$, a {\em bisection} $(X,Y)$ is a partition of $V$ into sets $X$ and $Y$ such that $|X|\le |Y|\le |X|+1$. The {\em size} of $(X,Y)$ is the number of edges
between $X$ and $Y$. In the {\sc Max Bisection} problem we are given a graph $G=(V,E)$ and are required to find a bisection of maximum size.
Corollary \ref{cor:1} in the next section shows that $\lceil m/2 \rceil$ is a tight lower bound on the maximum size of a bisection of $G$, where $m=|E|.$
In what follows, for any pair $U,W$ of disjoint sets of $V$, $(U,W)$ will denote the set of edges between $U$ and $W$, and $n$ and $m$ will stand for the number of vertices and edges, respectively, in the graph $G$ under consideration. In the rest of the paper, $n$ is assumed to be even as if $n$ is odd, we may add an isolated vertex to $G$ without changing the maximum size of a bisection or our lower bound of $\lceil m/2 \rceil$.

The standard parametrization of {\sc Max Bisection} is to decide whether $G$ has a bisection of size at least $k$. (We give basic definitions on parameterized complexity later in this section.) Using the $\lceil m/2 \rceil$ lower bound, it is easy to see that the standard parametrization of {\sc Max Bisection} has a kernel with at most $2k$ edges.
Indeed, if $\lceil m/2 \rceil\ge k$ the answer is {\sc yes} and, otherwise, $m\le 2k$. At the first glance, it looks like the size $2k$ of this kernel is small, but it is not true. Indeed, for $k>(m+1)/2$, we have $2k>m+1$, which means that the kernel is of little value from both theoretical and practical points of view.

Similar examples were given by Mahajan et al. \cite{MahajanRamanSikdar09} who indicated that only parameterizations above tight lower bounds or below tight upper bounds are of interest.
Several results on problems parameterized above tight lower bounds have already been obtained in the literature (e.g., \cite{AlonEtAl2009a,BollScott,CrowstonIPL,CrowstonSWAT,GutinKimLampisMitsou,GutinKimMnichYeo,GutinKimSzeiderYeo09a,GutinRafieySzeiderYeo07,GutinSzeiderYeo08,MahajanRaman99,MahajanRamanSikdar09,VillangerHeggernesPaulTelle}), but almost all results on the topic in the last couple of years were on constraint satisfaction rather than graph theoretical problems.

In this paper, we turn to graph theoretical problems parameterized above tight lower bounds and consider the following {\sc Max Bisection above Tight Lower Bound (Max-Bisec-ATLB)} problem: decide whether a graph $G$ has a bisection of size at least $\lceil m/2 \rceil+k,$ where $k$ is the parameter. We prove that this parameterized problem has a kernel with $O(k^2)$ vertices and $O(k^3)$ edges. Thus, in particular, {\sc Max-Bisec-ATLB} is fixed-parameter tractable.
A closely related result to ours is by Bollob{\'a}s and Scott \cite{BollScott} who proved that the problem of deciding whether a graph $G$ has a maximum cut of size at last $\frac{m}{2}+\sqrt{\frac{m}{8}}+k$, where $k$ is the parameter, has an algorithm of running time $O(2^{O(k^4)}+n+m)$, i.e., the problem is fixed-parameter tractable. Note that the problem considered by Bollob{\'a}s and Scott \cite{BollScott} is parameterized above a tight lower bound as $\lceil \frac{m}{2}+\sqrt{\frac{m}{8}+\frac{1}{64}} - \frac{1}{8}\rceil$ is a tight lower bound on the maximum size of a cut, which was first proved by Edwards \cite{Edwards}.

A \emph{parameterized problem} is a subset $L\subseteq \Sigma^* \times
\mathbb{N}$ over a finite alphabet $\Sigma$. $L$ is
\emph{fixed-parameter tractable} if the membership of an instance
$(x,k)$ in $\Sigma^* \times \mathbb{N}$ can be decided in time
$f(k)|x|^{O(1)},$ where $f$ is a computable function of the
parameter $k$. If the
nonparameterized version of $L$ (where $k$ is just a part of the input)
is $\NP$-hard, then the function $f(k)$ must be superpolynomial
provided $\P\neq \NP$. Often $f(k)$ is ``moderately exponential,''
which makes the problem practically feasible for small values of~$k$.
Thus, it is important to parameterize a problem in such a way that the
instances with small values of $k$ are of real interest.

Given a parameterized problem $L$,
a \emph{kernelization of $L$} is a polynomial-time
algorithm that maps an instance $(x,k)$ to an instance $(x',k')$ (the
\emph{kernel}) such that (i)~$(x,k)\in L$ if and only if
$(x',k')\in L$, (ii)~ $k'\leq g(k)$, and (iii)~$|x'|\leq h(k)$ for some
functions $g$ and $h$. The function $h(k)$ is called the {\em size} of the kernel.

It is well-known that
a parameterized problem $L$ is fixed-parameter
tractable if and only if it is decidable and admits a
kernelization. Due to applications, low degree polynomial size kernels are of main interest. Unfortunately, many fixed-parameter tractable problems do not have
kernels of polynomial size unless the polynomial hierarchy collapses to the third level \cite{BodlaenderEtAl2009a,BodlaenderEtAl2009,Fernau2009}.
For further background and terminology on parameterized complexity we
refer the reader to the monographs~\cite{DowneyFellows99,FlumGrohe06,Niedermeier06}.

\section{Results}\label{sec2}

A result similar to the following lemma but for cuts rather than bisections was apparently first proved by Haglin and Venkatesan \cite{HaglinVenkatesan}.

\begin{lemma}\label{lem:1}
If $M$ is a matching in a graph $G$, then $G$ has a bisection of size at least $\lceil m/2 \rceil + \lfloor |M|/2 \rfloor$.
\end{lemma}
\begin{proof}
Recall that we may assume that $n$ is even and let $p=n/2$. Let $U=u_1,u_2,\ldots ,u_p$ and $V=v_1,v_2,\ldots ,v_p$ be two disjoint sequences
of vertices of $G$ such that  $M = \{u_1v_1,\ldots,u_{|M|}v_{|M|}\}$. Starting from empty sets $X$ and $Y$, for each $i=1,2,\ldots ,p$,  place $u_i$ in $X$ or $Y$
with probability $1/2$ and place $v_i$ in the other set. Observe that the expectation of the size of the bisection is $|M|+(m-|M|)/2$ since the probability of each edge of $M$ to be between $X$ and $Y$ is 1 and the probability of any other edge to be between $X$ and $Y$ is $1/2$. Thus, there is a bisection in $G$ of size at least $\lceil m/2 + |M|/2\rceil\ge \lceil m/2 \rceil + \lfloor |M|/2 \rfloor.$

We can find such a bisection by derandomizing the above randomized procedure using the well-known method of conditional probabilities, see, e.g., Chapter 15 in \cite{alon} or Chapter 26 in \cite{jukna}.
This derandomization leads to a greedy algorithm in which at Step $i$ ($1\le i\le p$)  we place $u_i$ in $X$ and $v_i$ in $Y$ rather than the other way around if and only if $|(u_i,Y)|+|(v_i,X)|\ge |(u_i,X)|+|(v_i,Y)|,$ where $X$ and $Y$ are sets constructed before Step $i$ (here $u_i$ stands for $\{u_i\}$, etc.). The greedy algorithm takes time $O(m+n)$.
\end{proof}

\begin{corollary}\label{cor:1}
A graph $G$ has a bisection of size at least $\lceil m/2 \rceil$ and this lower bound on the maximum size of a bisection is tight.
\end{corollary}
\begin{proof}
The first part of the claim follows immediately from Lemma \ref{lem:1}. To see that $\lceil m/2 \rceil$ is tight, it suffices to consider the star $K_{1,m}$ for any odd $m$.
\end{proof}

\begin{theorem}
The problem {\sc Max-Bisec-ATLB} has a kernel with $O(k^2)$ vertices and $O(k^3)$ edges.
\end{theorem}
\begin{proof}
Recall that we may assume that $n$ is even, as otherwise we can add an isolated vertex. 
Let $M$ be a maximal matching in a graph $G=(V,E)$. Such a matching can be found in time $O(n+m)$. If $|M|\ge 2k$, then by Lemma \ref{lem:1}, the answer to {\sc Max-Bisec-ATLB} is {\sc yes}. Thus, assume that $|M|<2k.$
For each vertex $x$ covered by $M$, let $S(x)$ be the smallest of the following two sets: $N(x) \setminus V(M)$ and $V \setminus (V(M) \cup N(x))$, where $N(x)$ is the set of neighbors of $x$ and $V(M)$ is the set of vertices covered by $M.$ Now consider two cases.\\

\noindent{\bf Case 1:} There is a vertex $z\in V(M)$ with $|S(z)| \geq 2k -(|M|-1)$.
Let $X'$ be any set of size $2k-(|M|-1)$ in $N(z) \setminus V(M)$ and let $Y'$ contain $z$ and $2k-(|M|-1)-1$ vertices from $V(G) \setminus (V(M) \cup N(z))$.
Note that $|X'|=|Y'|=2k-|M|+1$ and there are $2k-|M|+1$ edges between
$X'$ and $Y'$. Furthermore $X'$  and $Y'$ are independent sets of vertices. Set $X=X'$ and $Y=Y'$, and let $M'$ be the set
of edges in $M$ minus the edge incident to $z$.
For each edge $uv$ in $M'$, place $u$ in $X$ or $Y$
with probability $1/2$ and place $v$ in the other set. Partition the vertices of $G$ still not in $X\cup Y$ into pairs
and use the randomized procedure of Lemma \ref{lem:1} to assign those vertices
to either $X$ or $Y$.

Observe that the expected number of edges between $X$ and $Y$ equals $|(X',Y')|+|M'|+f/2$, where $f$ is the number of edges of $G$ not belonging to $(X',Y')$ or $M'$. Thus, the expected number of edges between $X$ and $Y$ is at least $$m/2+[(2k-|M|+1) + (|M|-1)]/2=m/2+k.$$ Similarly to Lemma \ref{lem:1}, we can derandomize the randomized procedure
from the first paragraph of this proof to obtain a greedy-type algorithm producing a bisection of size at least $\lceil m/2 \rceil+k$.\\

\noindent{\bf Case 2:} $|S(x)| < 2k-|M|+1$ for all $x \in V(M)$. We start by performing the following reduction:
If $G$ has an independent set $I$ of size $n/2 +j$ (with $j>0$) such that all vertices in $I$ have the same neighborhood (and $I$ is maximal with respect to the two properties), then
delete $2j$ of the vertices in $I$ from $G$.  We may do this reduction as any bisection of $G$ will have at least $j$ vertices from $I$ in each part.
Note that if the reduction is performed, the new graph $G$ cannot have an independent set $I$ of size $n/2 +j$ (with $j>0$) such that all vertices in $I$ have the same neighborhood
(here $n:=n-2j$).

Now we will prove that $n=O(k^2).$
Let $S = \cup_{x \in V(M)} S(x)$ and note that $|S| \leq
2|M| (2k-|M|)$. Let $Z= V(G) \setminus (V(M) \cup S)$ and note that $|Z| \geq n - 2|M| - 2|M|(2k-|M|)$.
The  maximum value of the function $f(t) = 2t(2k-t+1)$ is obtained when $t=k+1/2$. However, for integral $t$, it is obtained when $t=k$ or $t=k+1$, which in both cases gives
$f(k)=f(k+1)=2k(k+1)$. Therefore $|Z| \geq n - 2k(k+1)$.
As all vertices in $Z$ have exactly the same neighborhood, we have $|Z| \leq n/2$, and thus we have the following:
$n/2 \geq |Z| \geq n - 2k(k+1)$ implying $4k(k+1) \geq n$.

Hence, we have a kernel with at most $4k(k+1)= O(k^2)$ vertices. Recall that $|M| < 2k$ and observe that
$V(G) \setminus V(M)$ is independent. Thus, the number of edges in the kernel is at most
$$|V(M)| \cdot |V(G) \setminus V(M)|  + {|V(M)| \choose 2}\le 4kn + 8k^2 = O(k^3).$$
\end{proof}

\section{Open Problems}

We have proved that {\sc Max-Bisec-ATLB} has a kernel with $O(k^2)$ vertices and $O(k^3)$ edges. It would be interesting to obtain a kernel with fewer vertices and/or edges.

We can obtain a stronger lower bound for the maximum size of a bisection in a graph $G=(V,E)$. Choose a random bisection $(X,Y)$ in $G$ by randomly choosing $n/2$ vertices of $G$.
Observe that the probability $p$ of an edge being in $(X,Y)$ is $\frac{n}{2(n-1)}$. Thus, $\lceil pm \rceil$
is a lower bound on the maximum size of a bisection. Observe that this bound is tight an the extreme graphs include not only stars, but also complete graphs.
It would interesting to determine the parameterized complexity of the following problem: given a graph $G$, decide whether $G$ has a bisection of size at least $\lceil pm \rceil+k,$ where $k$ is the parameter.

The situation between our main result and the last open problem is similar to that between the above-mentioned result of Bollob{\'a}s and Scott and
the following open question from \cite{MahajanRamanSikdar09}.
Determine the parameterized complexity of the following problem: given a connected graph $G$, decide whether $G$ has a cut of size at least $\frac{m}{2}+\frac{n-1}{4}+k,$ where $k$ is the parameter.
Note that $\frac{m}{2}+\frac{n-1}{4}$ is a tight lower bound on the maximum size of a cut of a connected graph, which was first proved by Edwards \cite{Edwards}.
It is easy to check that $\frac{m}{2}+\sqrt{\frac{m}{8}+\frac{1}{64}} - \frac{1}{8}\le \frac{m}{2}+\frac{n-1}{4}$.

\medskip

\paragraph{Acknowledgments}
Research of both authors was supported in part by an EPSRC
grant. Research of Gutin was also supported
in part by the IST Programme of the European Community, under the
PASCAL 2 Network of Excellence.

\urlstyle{rm}


\begin{thebibliography}{10}




 \bibitem{AlonEtAl2009a}
N.~Alon, G.~Gutin, E.~J. Kim, S.~Szeider, and A.~Yeo.
\newblock Solving {MAX}-$r$-{SAT} above a tight lower bound.
\newblock {\em Proc. SODA 2010}, 511--517.

\bibitem{alon} N. Alon and J. Spencer, The Probabilistic Method, 2nd Edition, Wiley, 2000.


\bibitem{BollScott} B. Bollob{\'a}s and A.D. Scott, Better bounds for Max Cut, in Contemporary Combinatorics, B. Bollob{\'a}s, ed.,
Bolyai Society Mathematical Studies 10:185--246, 2002.

\bibitem{BodlaenderEtAl2009a} H.~L. Bodlaender, R.G. Downey, M.R. Fellows,  and D. Hermelin. On problems without polynomial kernels. J. Comput. Syst. Sci. 75(8):423--434, 2009.

\bibitem{BodlaenderEtAl2009}
H.~L. Bodlaender, S.~Thomass{\'e}, and A.~Yeo.
\newblock Kernel bounds for disjoint cycles and disjoint paths.
{\em Proc. ESA 2009}, Lect. Notes
Comput. Sci. 5757:635--646, 2009.


\bibitem{CrowstonIPL}
R.~Crowston, G.~Gutin, and M.~Jones.
\newblock Note on {M}ax {L}in-2 above average.
\newblock Inform. Proc. Lett.  110:451--454, 2010.

\bibitem{CrowstonSWAT}
R.~Crowston, G.~Gutin, M.~Jones, E.~J. Kim, and I.~Ruzsa.
\newblock Systems of linear equations over $\mathbb{F}_2$ and problems
  parameterized above average.
\newblock {\em Proc. SWAT 2010}, to appear.

\bibitem{DowneyFellows99}
R.~G. Downey and M.~R. Fellows.
Parameterized Complexity,
Springer, 1999.

\bibitem{Edwards} C.S. Edwards, Some expremal properties of bipartite subgraphs. Canad. J. Math. 25: 475--485, 1973.

\bibitem{Fernau2009} H. Fernau, F.V. Fomin, D. Lokshtanov, D. Raible, S. Saurabh, and Y. Villanger, Kernel(s) for Problems with No Kernel: On Out-Trees with Many Leaves.
{\em Proc. STACS 2009}, 421--432.

\bibitem{FlumGrohe06}
J.~Flum and M.~Grohe.
Parameterized Complexity Theory,
Springer, 2006.

\bibitem{GutinKimLampisMitsou}
G. Gutin, E.J. Kim, M. Lampis, and V. Mitsou, Vertex Cover Problem Parameterized Above and Below Tight Bounds.
Theory Comput. Syst., to appear.

\bibitem{GutinKimMnichYeo} G.~Gutin, E.~J. Kim, M. Mnich, and
A.~Yeo.  Betweenness parameterized
above tight lower bound. J. Comput. Syst. Sci., in press.

\bibitem{GutinKimSzeiderYeo09a} G.~Gutin, E.~J. Kim, S.~Szeider, and
  A.~Yeo.  A probabilistic approach to problems parameterized
  above tight lower bound.   {\em Proc. IWPEC'09}, Lect. Notes
Comput. Sci. 5917:234--245, 2009.

\bibitem{GutinRafieySzeiderYeo07}
G.~Gutin, A.~Rafiey, S.~Szeider, and A.~Yeo.
The linear arrangement problem parameterized above guaranteed value.
Theory Comput. Syst., 41:521--538, 2007.

\bibitem{GutinSzeiderYeo08}
G.~Gutin, S.~Szeider, and A.~Yeo.
 Fixed-parameter complexity of minimum profile problems.
Algorithmica, 52(2):133--152, 2008.

\bibitem{HaglinVenkatesan} D.J. Haglin and S.M. Venkatesan, Approximation and intractability results for the
maximum cut problem and its variants, IEE Trans. Comput. 40(1):110--113, 1991.

\bibitem{jukna} S. Jukna,    Extremal Combinatorics: With Applications in Computer Science, Springer, 2001.

\bibitem{MahajanRaman99}
M.~Mahajan and V.~Raman.
Parameterizing above guaranteed values: {M}ax{S}at and {M}ax{C}ut.
J. Algorithms, 31(2):335--354, 1999.

\bibitem{MahajanRamanSikdar09}
M.~Mahajan, V.~Raman, and S.~Sikdar.
Parameterizing above or below guaranteed values.
 J. Computer System Sciences, 75(2):137--153, 2009.

\bibitem{Niedermeier06}
R.~Niedermeier.
Invitation to Fixed-Parameter Algorithms.
Oxford University Press, 2006.

\bibitem{VillangerHeggernesPaulTelle}
Y.~Villanger, P.~Heggernes, C.~Paul and J.~A. Telle.
Interval Completion Is Fixed Parameter Tractable.
SIAM J. Comput., 38(5):2007--2020, 2009.

\end{thebibliography}
\end{document}